\def\Here{}
\def\normalcolor{\rm }
\def\rmfamily{\rm }
\documentstyle[11pt,appb,epsf]{article}

\preprint{TPJU-13/96}\date{0}

\long\def\kom#1{}

\def\umod#1{\ifmmode #1\else $#1$\fi}

\def\alfs{\umod{\alpha_{\rm s}}}

\def\beqa{\begin{eqnarray}}
\def\beqal{\begin{eqlettarray}}
\def\beq{\begin{equation}}

\def\eeqa{\end{eqnarray}}
\def\eeqal{\end{eqlettarray}}
\def\eeq{\end{equation}}

\def\Eq#1{Eq.(\ref{#1})}

\def\ie{{\sl i.e. }}

\def\MB{M_{\rm B}}

\def\nn{\nonumber}

\def\P{{\cal P}}

\def\slash{\rlap/}
\def\sthWsq{\sin^2\theta_{\rm W}}

\def\Tr{\mathop{\rm Tr}}
\def\tthWsq{\tan^2\theta_{\rm W}}
\def\tthW{\tan\theta_{\rm W}}
\def\vex#1{\vbox{\kern#1 pt}}

\def \J{{\cal J}}

\def \MA{M_{\rm A}}
\def \MB{M_{\rm B}}
\def\rW{\rho_{\rm W}}
\def\elam{{\rm e}^-_\lambda}

\newlength{\figwidth}
\def\figscale{0.62}

\begin{document}
\eqsec
\title{ON THE ELECTRON STRUCTURE FUNCTION%
\thanks{Work supported by the Polish State Committee for Scientific Research
(grant No.~2~P03B~081~09) and the Volkswagen Foundation.}
}
\author{Wojciech S\L{}OMI\'NSKI$^A$ and Jerzy SZWED$^{A,B}$%
 \thanks{J. William Fulbright Scholar.}
\address{
$^A$Institute of Computer Science, Jagellonian University,
Reymonta 4, 30-059~Krak\'ow, Poland\\
$^B$Physics Department, Brookhaven National Laboratory,\\ 
Upton, NY 11973, USA}
}

\maketitle
\begin{abstract}
The collinear QCD structure of the electron is studied within the
Standard Model. The electron structure function is defined and calculated in 
leading logarithmic approximation. It shows important contribution from 
the interference of the intermediate electroweak bosons. The problem of 
momentum scales is extensively discussed. The master equations for
the QCD parton densities inside the electron are constructed and solved
numerically in the asymptotic region. Significant 
corrections to the naive evolution
procedure are found. 
Phenomenological applications at present
and future momentum scales are discussed.
\end{abstract}
  
\bigskip
\centerline{\sl Dedicated to Andrzej Bia\l{}as in honour of his 60$^{\sl th}$
birthday}


\section{Introduction}
It is the nature of the Quantum Chromodynamics (QCD) and its running coupling 
constant which cause point-like objects of the electroweak theory to
develop quark and gluon structure. This QCD component was first calculated
\cite{resph}  in the case of the photon.
In the forward direction, when (nearly) massless quarks are emitted 
collinearly from the photon, a consecutive QCD cascade  developes practically 
`at no cost'. The presence of such a `resolved' photon has been observed
in many experiments \cite{resphexp} allowing now for a precise study
of its dependence on the momentum fraction \(x\) and the momentum scale $P^2$. 
Analogous calculation  extended to the
case of the weak bosons W and Z has also been performed \cite{WS1}. It shows
several differences as compared to the photon case. Due to the nature of  
weak couplings the QCD parton densities depend
strongly on the quark flavour and in most cases the polarized densities
turn out to be nonzero. The longitudinal bosons show much weaker structure
and in the leading logarithmic approximation can be neglected. 
Phenomenological 
applications of the `resolved' W and Z bosons require momentum scales much higher
than their masses.
This is because
only in this region the approximations used 
are valid, moreover, only there the photon contribution does not 
dominate 
entirely the weak bosons' effects.

In order to 
study the boson structure functions we need  a source of real electroweak
bosons. But even in the case of the photon, `real' means in most cases
only `nearly on-shell'. The situation is even more approximate with the
weak bosons. Here `nearly real' means in practice that their  momentum squared
is negative and close to zero. This is because
the best known source of  high energy electroweak
bosons is a leptonic beam. The procedure applied usually
is to replace the incoming lepton with the spectrum of equivalent
bosons \cite{WW,eqb}, $F^{e^-}_B$, 
and to convolute it with the parton distributions, $F^B_h$, inside the boson,
which results in the following QCD parton $h$ distribution inside the electron:
\beq
F^{e^-}_h(z,\hat Q^2,P^2)=\sum_{B=\gamma,\rm W,Z} \int dx\,dy\,\delta(z-xy) \,
F^B_h(x,P^2) F^{e^-}_B(y,\hat Q^2)\,,
\label{FF}
\eeq 
where
 $P^2$ is the 
hard process 
scale,
$\hat Q^2$ is the maximum allowed virtuality of the boson (usually taken 
to be proportional to $P^2$) and $z$ is the momentum fraction of the parton $h$ with respect to the
electron (detailed definitions follow). 

The procedure presented above has, however, several unclear points.
It does not answer the questions: 
what is the relation between the two entering 
momenta scales $\hat Q^2$ and $P^2$, 
how far off-shell can the electroweak bosons be, are there any
interference effects between the intermediate bosons (is the sum actually diagonal
in $B$ and/or in the boson polarizations). 
One may also ask the question: when is the convolution
\Eq{FF} justified and what are the corrections? The 
best way to clarify these doubts is to address a direct
question: what is the QCD content of the incoming electron? 
In this paper we
demonstrate how 
the concept of the 
electron structure function
allows for a detailed study of the QCD cascade originating from an 
electron.

The paper is organized as follows. In Section 2 we discuss the general
structure of the QCD master equations in the system of leptons
and electroweak bosons. The leading order calculation of splitting functions
of an electron into a quark (antiquark) is demonstrated in Section 3.
The problem of momentum scales and interference 
effects are discussed there in detail.
The constructed master equations for the parton densities
inside the electron are solved in the asymptotic region in Section 4. 
We demonstrate
there the flavour and spin dependence of the asymptotic distributions.
Comments on phenomenological applications, 
summary and conclusions close the paper.

\section{Master equations}
In order to study the QCD content of a lepton $\ell$ we look 
for partons
produced by a highly virtual
gluon G$^*$ scattering off $\ell$:
\begin{equation}
{\rm G}^* + \ell \rightarrow \ell' + h + {\rm anything}\,.
\label{reaction}
\end{equation}
where $h$ = quark q, antiquark ${\rm\bar q}$ or gluon G and $\ell'$ is the outgoing lepton.
We have chosen the gluon  G$^*$ 
as a probe because it does not couple directly
to the lepton. Our process, \Eq{reaction}, is a part of many
physical reactions, the far off-shell gluon may originate from 
an incoming hadron or another lepton.
The virtualness $P^2$ of this gluon  sets the scale for the process.

Before going to a detailed calculation 
we  present a heuristic derivation of the
master
equations for parton densities involved in the process (\ref{reaction}).
We  
 point out which assumptions usually made in this scheme
are not fullfilled in a general case.
The  standard master  equation for any `parton' ${\cal A}$
density inside $\ell$, $F^{\ell}_{\cal A}$, reads\cite{AP}
\beq
{d\over dt}F^{\ell}_{\cal A}(z,t) = \sum_{\cal B} \int dx\,dy\,\delta(z-xy)
\,\P_{\cal A}^{\cal B}(x,t)
\, F_{\cal B}^{\ell}(y,t)
\eeq
or, using a shorthand notation for the convolution
\beq
{d\over dt}F^{\ell}_{\cal A}(t) = \sum_{\cal B} 
\P_{\cal A}^{\cal B}(t)
 \otimes
 F_{\cal B}^{\ell}(t)
\,.
\label{meqgen}
\eeq
In general $t= \log(P^2/P_0^2)$, where $P_0$ is some momentum scale.
We discuss the choice of $P_0$ in the next section.
The `splitting function' $\P_{\cal A}^{\cal B}(x,t)$ is the probability
density per unit $t$ to find ${\cal A}$ carrying fraction $x$ of
${\cal B}$'s momentum in a collinear decay of ${\cal B}$.
The above master equations are quite general within the approximation that
the probability to produce a parton in a fusion process is negligible.
It is assumed here that this probabilistic approach holds for a physical
process \ie the contributions from different
`intermediate' partons ${\cal B}$ do not mix. 
As we will show in the next section, this is not the case for the
electroweak sector.

Since there is no direct lepton-gluon coupling the process (\ref{reaction}) must be
mediated by electroweak bosons and we have to consider following
types of 
`partons' inside lepton $\ell$: q, ${\rm\bar q}$, G, $\ell,\,\gamma$, W$^\pm$,
Z$^0$. 
Assuming for the moment that the probabilistic approach holds, 
\Eq{meqgen} takes the form
\beqa
{d\over dt}F^{\ell}_{\cal A}(t) &=&
\sum_{\ell'}
\P_{\cal A}^{\ell'}(t)
 \otimes
 F_{\ell'}^{\ell}(t)
+
\sum_{B=\gamma, \rm W, Z} 
\P_{\cal A}^{B}(t)
 \otimes
 F_{B}^{\ell}(t)\nn\\
&+&
\sum_{h={\rm q, \bar q, G}}
\P_{\cal A}^h(t)
 \otimes
 F_{h}^{\ell}(t)
\label{meq1}
\,.
\eeqa

Within the approximation of the  lowest order in the electromagnetic coupling 
$\alpha$ 
and leading-log order in the strong coupling $\alfs(t)$
\beqa
F_{\ell'}^{\ell}(x,t) &=& \delta_\ell^{\ell'}\, \delta(1-x) + {\cal O}(\alpha^2)
\,,\\
{\cal P}_B^\ell(x,t) &=& {\alpha\over 2\pi} P_B^\ell(x)
\,,\\
{\cal P}_h^B(x,t) &=& {\alpha\over 2\pi} P_h^B(x)
\,,\\
{\cal P}_{h'}^h(x,t) &=& {\alfs(t)\over 2\pi} P_{h'}^h(x)
\eeqa
and all other $\P_{\cal A}^{\cal B} = 0$.

Inserting above relations into \Eq{meq1} we arrive at
\begin{eqlettarray}
\eqalabel{meq}
{d\over dt}F^{\ell}_{B}(t) &=&
{\alpha\over 2\pi} P_B^\ell\,,
\label{meq-B}
\\
{d\over dt}F^{\ell}_{h}(t) &=&
{\alpha\over 2\pi} 
\sum_{B=\gamma, \rm W, Z} 
P_h^B
 \otimes
 F_{B}^{\ell}(t)
+
{\alfs(t)\over 2\pi} 
\sum_{h'={\rm q, \bar q, G}}
P_h^{h'}
 \otimes
 F_{h'}^{\ell}(t)
\label{meq-h}
\,.\;\;\;\;\;\;
\end{eqlettarray}

The solution to \Eq{meq-B} is known as the density of `equivalent bosons' in
the lepton $\ell$ \cite{eqb}. 
The splitting functions $P_h^B$ vanish for $h={\rm G}$.  $P_{\rm q}^B$
and $P_{\rm\bar q}^B$ have  been calculated for $B=\gamma$ in Ref. \cite{resph} and 
for $B={\rm W,Z}$ in Ref. \cite{WS1}. 
$P_h^{h'}$ are the standard Altarelli-Parisi splitting functions \cite{AP}.
Inserting equivalent boson density $F^{\ell}_{B}$ into \Eq{meq-h} one
can solve it for $F^{\ell}_{h}$.
Eqs.(\ref{meq}) suggest also
 that there is one common
momentum scale governing the whole system. This is not true in general
as will be demonstrated below.

As already stated 
the above reasoning neglects interference effects, here possible
between $\gamma$ and Z bosons. Because of different masses and different
couplings
of these bosons to leptons this interference is non-trivial, \ie
cannot be removed 
by any kind of `diagonalization' procedure\footnote{The Standard Model
gauge group is not simple as opposed to QCD where all colour
interferences are taken into account by a group-theoretical weight
factor independent of the couplings and kinematical variables.}.
Thus, instead of $F^{\ell}_{B}$,
one should rather use a density matrix, $F^{\ell}_{AB}$, of
equivalent bosons inside the lepton $\ell$. In the following section we
perform an explicit calculation which shows how this
density matrix arises and how to
generalize the first term of \Eq{meq-h} in order to take the $\gamma$-Z
interference into account.

The use of the density matrix implies --- at the first sight --- that
the probabilistic approach expressed in terms of the master equations
breaks down. If, however, instead of looking at
the contributions from different electroweak bosons
we introduce the notion of the hadronic content of the lepton,
the probabilistic interpretation is recovered and
the master equations read 

\beq
{d\over dt}F^{\ell}_{h}(t) =
{\alpha\over 2\pi} 
P_h^{\ell}(t)
+
{\alfs(t)\over 2\pi} 
\sum_{h'={\rm q, \bar q, G}}
P_h^{h'}
 \otimes
 F_{h'}^{\ell}(t)
\label{meq-e}
\,,
\eeq
where $P_h^{\ell}(t)$ is `lepton to hadron splitting function'. 
In the next Section we define and calculate this function.


\def\elam{{\rm e}^-_\lambda}
\section{QCD current of electron}
The hadronic structure of a lepton can be analyzed
by looking at any process where the lepton couples to the QCD current.
In the following we will calculate the matrix elements squared of
this current in the case of $\ell = {\rm e}^-$. 

In the lowest order in the electromagnetic and strong
coupling constants ($\alpha$ and $\alpha_{\rm s}$)
the current matrix element squared
for polarized particles reads:
\begin{eqnarray}
{\J}_{\mu\nu}^{\lambda\eta}(l,p) =
{1 \over 4 \pi}
\int  d\Gamma_{l'} d\Gamma_{k} d\Gamma_{k'} (2\pi)^4 \delta_4(k+k'+l'-p-l) 
&&\nonumber\\
\times
\sum_{\ell'={\rm e},\nu} \; \sum_{\lambda'=\pm}
 \langle \elam | J_\mu^\dagger(0)|\ell'_{\lambda'}\, {\rm q_{\eta}\, \bar q}_{-\eta}\rangle
\langle \ell'_{\lambda'}\, {\rm q_{\eta}\, \bar q}_{-\eta}|J_\nu(0)|\elam\rangle \,,
&&
\label{Jmunu}
\end{eqnarray}
where
\beq
 d\Gamma_{k} = {d^4k \over (2\pi)^4} 2\pi \delta(k^2)
\eeq 
and indeces $\lambda, \lambda',\eta$ denote helicities of corresponding
particles.
We do not sum over quark flavours and helicities
because we are interested in the densities of polarized partons in the
electron. In other words we choose a single flavour of particular
helicity in the final state.

\begin{figure}[h]
\centerline{
\mbox{\epsfxsize=0.9\columnwidth \epsfbox{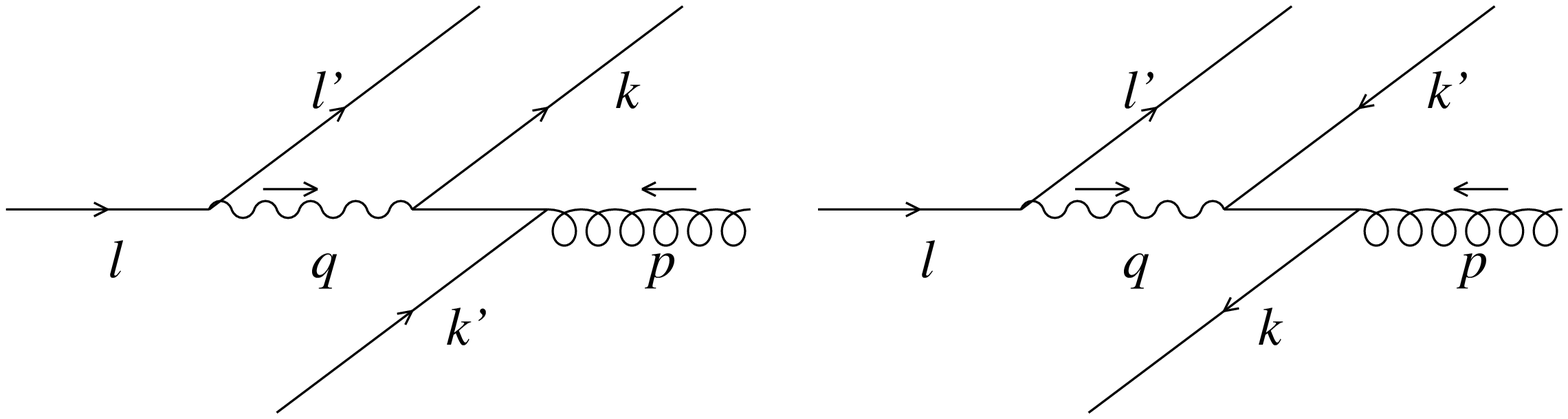}}
}
\caption{Lowest order graphs contributing to the process: $e + G^* \rightarrow \ell' + q + \bar q$.}
\label{F-graf}
\end{figure}

The Feynman graphs contributing to 
our process are shown in Figure\thinspace \ref{F-graf}. 
The incoming electron e$^-$ carries
4-momentum $l$ and the off-shell gluon G$^*$ of 4-momentum $p$ with
large $P^2 \equiv -p^2$, supplies the QCD current.
In the final state we have massless quark q and antiquark ${\rm\bar q}$
of 4-momenta $k$ and $k'$ 
and lepton $\ell'$ (electron or neutrino) of 4-momentum $l'$.
The lepton interacts with the quark exchanging an electroweak boson
$B = \gamma,{\rm Z,W}$ which carries  4-momentum $q$ ($Q^2 \equiv -q^2$).
For massless quarks ${\J}_{\mu\nu}^{\lambda\eta}(l,p)$ is known to
have only 3 independent components
for any given polarizations $\lambda, \eta$. 
They can be conveniently
paramatrized by the helicities of the incoming gluon.
Let us chose a reference frame where the gluon momentum is parallel to
the $z$-axis, $p^\mu =(p_0,0,0,p_z)$.
Its polarization vectors 
$e_{(\sigma)}^\mu(p)$ read
\begin{eqlettarray}
  e_{\pm}^\mu&=& {1\over\sqrt 2} (0,1,\pm i,0)\,,\\
  e_0^\mu(p)&=& {1\over P}\,(p_z,0,0,p_0)\,,
\end{eqlettarray}
where $P=\sqrt{|p^2|}$.

The three helicity components of $\J$
\begin{equation}
{\J}_{\sigma}^{\lambda\eta}(l,p) =
e_{(\sigma)}^{\mu *}(p)\, 
{\J}_{\mu\nu}^{\lambda\eta}(l,p)\, e_{(\sigma)}^\nu(p)
\end{equation}
are proportional to the cross-section for the (virtual) process
\begin{equation}
\elam  + {\rm G}_\sigma^* \rightarrow {\ell}'_{\lambda'} + {\rm q}_\eta
+ {\rm\bar q}_{-\eta} \,.
\end{equation}

We will proceed with the calculation of ${\J}_{\sigma}^{\lambda\eta}$ 
in the reference frame where the $B$ boson momentum $\vec q= -\vec p$, 
\ie in the center-of-mass of $B$ and G$^*$, which is also the rest frame
of the ${\rm q\bar q}$ pair.
We choose $z$-axis in the direction of $\vec q$, the transverse 
directions are measured then in the $xy$ plane. In this frame the  momenta 
have following components:
\begin{eqnarray}
q&=&(q_0,0,q_z)\,,\nn\\
p&=&(p_0,0,-q_z)\,,\nn\\
l&=&(l_0,\vec l_\perp,l_z)\,,\nn\\
l'&=&(l'_0,-\vec l_\perp,l'_z)\,,\nn\\
k&=&(k_0,\vec k_\perp,k_z)\,,\nn\\
k'&=&(k_0,-\vec k_\perp,-k_z)\,,
\end{eqnarray}
where we have used the fact that the quarks are massless.

\kom{
It is convenient to parametrize the momenta in term of the light-cone
components. To this end let us introduce light-like 4-vectors:
\begin{equation}
R={1\over\sqrt 2} (1,0,0,1)\; {\rm and}\;
L={1\over\sqrt 2} (1,0,0,-1)\,.
\end{equation}
We can decompose now all the momenta as follows
\begin{eqnarray}
q &=& q_{\rm R}R + q_{\rm L}L\,,\\
p &=& p_{\rm R}R + p_{\rm L}L\,,\\
l &=& l_{\rm R}R + l_{\rm L}L + l_\perp\,,\\
l' &=& l'_{\rm R}R + l'_{\rm L}L - l_\perp\,,\\
k &=& k_{\rm R}R + k_{\rm L}L + k_\perp\,,\\
k' &=& k_{\rm L}R + k_{\rm R}L - k_\perp\,.
\end{eqnarray}
}
Introducing the momentum of the q${\rm\bar q}$ pair,
$K=k+k'$ we can decompose the 3-body phase space of \Eq{Jmunu} as
follows
\begin{equation}
  d\Gamma_{l'} d\Gamma_{k} d\Gamma_{k'} (2\pi)^4 \delta_4(k+k'+l'-p-l) 
=
 {1\over 2\pi} dW^2 d\Phi_D d\Phi_2\,,
\end{equation}
where $W^2=K^2$ is the invariant mass squared of the ${\rm q\bar q}$
pair, $d\Phi_2$ is the 2-body phase space of $l + p \rightarrow l' + K$
and $d\Phi_D$ is the phase space of $K \rightarrow k+k'$ decay.
The explicit expressions read
\beq
d\Phi_D = {d\cos\omega \over 16\pi}\, {d\chi \over 2\pi}\,,
\eeq
where $\omega$ and $\chi$ are the polar and azimuthal angles of
$\vec k$ and
\beq
d\Phi_2 = {dQ^2 \over 8\pi \Lambda(s,p^2,m^2)}\, {d\varphi \over 2\pi}\,,
\eeq
where $\varphi$ is the azimuthal angle of $\vec l'$ 
The function
$\Lambda$ is defined as $\Lambda(a,b,c)=\sqrt{(a+b-c)^2-4ab}$
and $s$ is the total energy squared:
\beq
s = (p+l)^2\,.
\eeq

\def\cL{{\cal L}}
\def\cH{{\cal H}}

\begin{figure}[hbt]
\begin{center}
\mbox{\epsfxsize=3cm\epsfbox{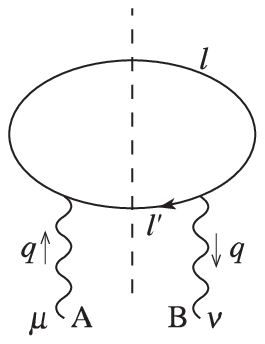}}
\caption{Lowest order diagram for $\cL^{\mu\nu}$.}
\label{f-lloop}
\end{center}
\end{figure}

\begin{figure}[hbt]
\begin{center}
\mbox{\epsfxsize=\columnwidth\epsfbox{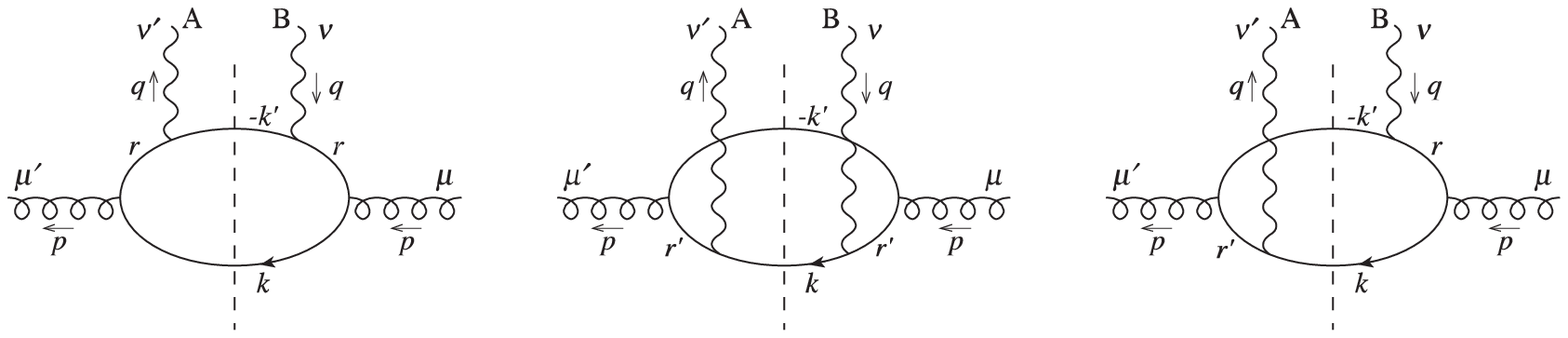}}
\caption{Lowest order diagrams for $\cH^{\nu'\nu}$.}
\label{f-hloop}
\end{center}
\end{figure}

For given final lepton $\ell'$ (e$^-$ or $\nu_{\rm e}$) the current
matrix element squared (without phase space factors) can be written as
\begin{equation}
\sum_{\scriptstyle A,B=\atop
\gamma, \rm W, Z}
 \cL^{\mu'\mu}(\lambda,\lambda';A,B)\, \cH^{\nu'\nu}(\sigma,\eta;A,B)
D_{\nu'\mu'}(A) D_{\nu\mu}(B)\,.
\label{mxel}
\end{equation}
Here, $D_{\mu\nu}$ denote the propagators of electro-weak bosons.
The leptonic part
$\cL^{\mu\nu}(\lambda,\lambda';A,B)$ is given by the Feynman diagram
of Fig.\thinspace\ref{f-lloop} and the hadronic one reads
\beq 
\cH^{\nu'\nu}(\sigma,\eta;A,B) =
e_{(\sigma)}^{*\mu'}(p)\, 
\cH^{\nu'\nu}_{\mu'\mu}(\eta;A,B)
\, e_{(\sigma)}^\mu(p)
\eeq
with  $\cH^{\nu'\nu}_{\mu'\mu}(\eta;A,B)$ given by the sum of diagrams
shown in Fig.\thinspace\ref{f-hloop} for the outgoing quark of helicty $\eta$.

For massless quarks the hadronic tensor satisfies
\beq 
q^\mu\cH_{\mu\nu}(\sigma,\eta;A,B) = 
q^\nu\cH_{\mu\nu}(\sigma,\eta;A,B)=0
\label{qH}
\eeq
which allows us to take
\begin{equation}
D_{\mu\nu}(A) = {1\over Q^2+\MA^2}\, g_{\mu\nu}\,.
\label{prop}
\end{equation}

These propagators can be expressed by the polarization vectors
by means of the
following identity:
\begin{equation}
e^{*\mu}_+(q)\,e^\nu_+(q)
+ e^{*\mu}_-(q)\,e^\nu_-(q)
+ {q^2 \over |q^2|}  e^{*\mu}_0(q)\,e^\nu_0(q)
= {q^\mu q^\nu \over |q^2|} - g^{\mu\nu}\,.
\end{equation}

\def\Lep{L}
\def\Had{H}
Using again equations (\ref{qH})
we find that there is no interference between different polarizations of
the exchanged bosons.
Thus we obtain the formula for
${\J}_{\sigma}^{\lambda\eta}$  where the contributions from the leptonic
and hadronic subprocesses are clearly separated:
\begin{eqnarray}
{\J}_{\sigma}^{\lambda\eta} &=&
 {1\over 2^{10}\pi^4} {1 \over  \Lambda(s,p^2,m^2)}\,
\sum_{\scriptstyle A,B=\atop \scriptstyle \gamma, \rm W, Z}
\int
{dQ^2\, dW^2 \over (Q^2+\MA^2)(Q^2+\MB^2)}\,
\nn\\
&\times&
\sum_{\rho}
 \Lep_{\lambda\rho}(A,B)
\; \Had_{\sigma\eta\rho}(A,B)
\,,
\label{JLH}
\end{eqnarray}
where
\begin{eqnarray}
 \Lep_{\lambda\rho}(A,B)&=&\sum_{\lambda'}\int {d\varphi \over 2\pi}\,
e_{(\rho)\mu}(q)\, 
 \cL^{\mu\nu}(\lambda,\lambda';A,B) 
 e^*_{(\rho)\nu}(q)\,,
\\
\Had_{\sigma\eta\rho}(A,B) &=& \int {d\chi \over 2\pi}\,{d\cos\omega }\,
e^*_{(\rho)\mu}(q)\, 
 \cH^{\mu\nu}(\sigma,\eta;A,B)
 e_{(\rho)\nu}(q)\,.\;\;\;\;\;
\label{Hhel}
\end{eqnarray}

The calculation of the leptonic part,
$\Lep_{\lambda\rho}(A,B)$ is straightforward and 
the result is given by
\beqa
\Lep_{\lambda\rho}(A,B)&=&\sum_{\lambda'}\int\! {d\varphi \over 2\pi}\,
 g^{A\ell}_{\lambda'} g^{B\ell}_\lambda 
\left\{
\Tr [\slash l \slash e_{(\rho)} P_{\lambda'} \slash l' \slash e^*_{(\rho)} P_\lambda]
+mm'\Tr [\slash e_{(\rho)}  P_{\lambda'}\slash e^*_{(\rho)}  P_\lambda]
\right\}\nn\\
&\equiv&\sum_{\lambda'}
 g^{A\ell}_{\lambda'} g^{B\ell}_\lambda 
L_{\lambda'\lambda\rho}
\,,
\label{lmm}
\eeqa
where $e_{(\rho)}\equiv e_{(\rho)}(q)$,
$g^{B\ell}_\lambda$ denotes the coupling of the electro-weak boson
$B$ to the lepton $\ell$ of helicity $\lambda$, $m$ and $m'$ are the
masses of initial and final lepton, respectively, and 
$P_\lambda$ are projection operators:
\begin{equation}
P_\pm = {1\over 2} (1 \pm\gamma_5)\,.
\end{equation}

\kom{
The result for $\cL$ can be expressed in terms of projections on the
boson helicity states.
Performing the integration over the azimuthal angle $d\varphi$ 
we can define
\begin{equation}
L_{\lambda'\lambda\rho} = 
\int {d\varphi \over 2\pi}\,
e_{(\rho)\mu}(q)\, \bar L^{\mu\nu}_{\lambda\lambda'}\,
 e^*_{(\rho)\nu}(q)\,.
\end{equation}
All elements non-diagonal in boson helicities vanish.
}

The explicit expressions for $L_{\lambda'\lambda\rho}$ read
\beqa
   L_{\rm LL-} =   L_{\rm RR+} &=& 2 Q^2 Y_+(y) \,,\\
   L_{\rm RR-} =   L_{\rm LL+} &=& 2 Q^2 Y_-(y) \,,\\
   L_{\rm LL0} =   L_{\rm RR0} &=& 2 Q^2 Y_0(y)  - 2 mm' \,,\\
   L_{\rm LR+} =   L_{\rm RL+} =   L_{\rm LR-} =   L_{\rm RL-} &=& -2 mm' \,,\\
   L_{\rm LR0} =   L_{\rm RL0} &=&  2 mm' \,,
\eeqa
where 
\beqa
Y_+(y) &=& {1 \over y^2}\,,\\
Y_-(y) &=& {(1-y)^2 \over y^2}\,,\\
Y_0(y) &=& 2{1- y \over y^2}
\eeqa
and $y$ is defined as 
\beq
y = {q_0 + q_z \over l_0 + l_z}\,.
\label{ydef}
\eeq

In the case of large energy scales we can neglect the electron mass $m$
which results in $L_{\lambda'\lambda\rho}$ being diagonal in the lepton
helicities. Thus for $\Lep_{\lambda\rho}$, \Eq{lmm}, we obtain
\beqa
\Lep_{\lambda\rho}(y,Q^2;A,B) &=&
4\pi\alpha
 \hat g^{A\ell}_{\lambda}
 \hat g^{B\ell}_\lambda 
L_{\lambda\lambda\rho}(y,Q^2)\nn\\
&=&
8Q^2\pi\alpha
  \hat g^{A\ell}_\lambda \hat g^{B\ell}_\lambda
\left\{
\begin{array}{ll}
Y_{+}(y) &{\rm for}\;\rho=\lambda \\ 
Y_{-}(y) &{\rm for}\;\rho=-\lambda \\ 
Y_{0}(y) &{\rm for}\;\rho=0
\end{array} 
\right.
\,.
\label{Lres}
\eeqa
We have factored out the explicit dependence on $\alpha$ by defining the
coupling constants scaled by the positron
charge $\hat g^{A\ell}_\lambda = g^{A\ell}_\lambda / e$.

%
%

Let us proceed now with
the hadronic part. 
First we calculate the integrand of \Eq{Hhel}.
It is given by the Feynman graphs of Fig.\thinspace\ref{f-hloop}:
\begin{equation}
e^*_{(\rho)\mu}(q)\, 
 \cH^{\mu\nu}(\sigma,\eta;A,B)
 e_{(\rho)\nu}(q)
=
G_{11} + G_{22} + 2{\rm Re} G_{12}\,.
\label{Hgraf}
\end{equation}
Explicit expressions for $G_{ik}$ read
\beqa
 G_{11} &=&
-{f}_{\rm c}\;
 g_{\rm s}^2 g^{Aq}_\eta g^{Bq'}_\eta
 {1\over r^4}
 \Tr[\slash k'
     \slash e^*_{(\rho)}(q)
     \slash r \slash e^*_{(\rho)}(p)
     P_\eta \slash k \slash e_{(\rho)}(p)
     \slash r
     \slash e_{(\rho)}(q)
      ]\,,\nn\\
 G_{22} &=&
-{f}_{\rm c}\;
 g_{\rm s}^2 g^{Aq}_\eta g^{Bq'}_\eta
 {1\over r'^4}
 \Tr[\slash k'
     \slash e^*_{(\rho)}(p)
     \slash r'
     \slash e^*_{(\rho)}(q)
     P_\eta \slash k
     \slash e_{(\rho)}(q)
     \slash r'
     \slash e_{(\rho)}(p)
      ]\,,\nn\\
 G_{12} &=&
-{f}_{\rm c}\;
 g_{\rm s}^2 g^{Aq}_\eta g^{Bq'}_\eta
 {1\over r^2 r'^2}
 \Tr[\slash k'
     \slash e^*_{(\rho)}(p)
     \slash r'
     \slash e^*_{(\rho)}(q)
     P_\eta \slash k
     \slash e_{(\rho)}(p)
     \slash r
     \slash e_{(\rho)}(q)
      ]\,,
\;\;\;\;\;\;\;\;\;\;\;
\label{Htraces}
\eeqa
where ${f}_{\rm c}$ is the colour factor, 
$g_{\rm s}$ --- the strong coupling constant and
\begin{equation}
r = k - p \,,
\;\;\;
r' = p - k'\,.
\end{equation}

${f}_{\rm c}$ is the same for all diagrams and independent
of the weak boson type. 
In our case (SU($N$) colour group and fixed final quark flavour)
\beq
{f}_{\rm c} = {1\over 2}\,.
\eeq

On substituting $P_\eta = (1 +\eta\gamma_5)/2$ into \Eq{Htraces}
we see that \Eq{Hgraf}
becomes the sum of two parts --- with and without the $\gamma_5$ matrix
under the traces.
In order to perform the integration of \Eq{Hhel}
we notice that
the denominators of quark propagators depend only on $\omega$ and read
\beqa
r^2 = (k-p)^2 &=& p^2 - 2pk 
= -{W^2+Q^2+P^2 + \Lambda(W^2,q^2,p^2) \cos\omega \over 2}\,,\nn\\
r'^2 = (k'-p)^2 &=& p^2 - 2pk' 
= -{W^2+Q^2+P^2 - \Lambda(W^2,q^2,p^2) \cos\omega \over 2}\nn\,.
\eeqa

The `axial' part (containing $\gamma_5$) vanishes upon integration
over the angles and we obtain the following result for the hadronic part
\beq
H_{\rho\sigma\eta} =
8{f}_{\rm c}\;
 g_{\rm s}^2 g^{Aq}_\eta g^{Bq'}_\eta
h_{\rho\sigma}(x,\tau) 
\,,
\label{Hres}
\eeq
which depends on two dimensionless kinematic variables $x$ and $\tau$:
\beq
x = -{p_0+p_z\over q_0+q_z}
\label{xdef}
\eeq
and
\beq
\tau = 1 - {\Lambda(W^2,q^2,p^2)\over W^2 + P^2 + Q^2}\,.
\label{tau}
\eeq

The explicit expressions for $h_{\rho\sigma}(x,\tau)$ 
are given in Appendix A.
In the following considerations
we will need only their leading terms for $P^2 \gg Q^2$.

Before quoting these approximate results let us look at the 
exact formula for the current matrix element squared, \Eq{JLH}.
Substituting the results from \Eq{Lres} and \Eq{Hres} we get
\begin{eqnarray}
{\J}_{\sigma}^{\lambda\eta} &=&
 {\alpha^2\alfs \over 2\pi} 
\int_0^{W^2_{\rm max}} {dW^2  \over  \Lambda(s,p^2,m^2)}\,
\sum_{\scriptstyle A,B=\atop \scriptstyle \gamma, \rm W, Z}
\int_{Q^2_{\rm min}}^{Q^2_{\rm max}}
{Q^2 dQ^2 \over (Q^2+\MA^2)(Q^2+\MB^2)}\,
\nn\\
&\times&
\sum_{\rho}
 {1\over y} P^{\ell_\lambda}_{A_\rho B_\rho}(y)
\; 
\hat g^{Aq}_\eta \hat g^{Bq'}_\eta
h_{\rho\sigma}(x,\tau)
\,,
\end{eqnarray}
where, by definition,
\beq
P^{\ell_\lambda}_{A_\rho B_\rho}(y)
\equiv
{y\over 8\pi\alpha Q^2}  \Lep_{\lambda\rho}(y,Q^2;A,B)\,.
\label{Pdef}
\eeq
$\tau, x, y$ can be expressed in terms of the integration
variables, $W^2$ and $Q^2$. $\tau$ is given by \Eq{tau} while
$x$ and $y$, defined by Eqs. (\ref{xdef}) and (\ref{ydef}), read
\begin{eqnarray}
x &=& 
 {\Lambda + P^2 - Q^2 -W^2
\over 
  \Lambda + P^2 -Q^2 + W^2 
}\,,\\
y &=& 
{2\Lambda
\over 
2(s - m^2) + \Lambda +P^2 -Q^2 -W^2
}\,,
\end{eqnarray}
where $\Lambda\equiv \Lambda(W^2,q^2,p^2)$.

The integration limits defined by kinematics read:
\beqa
W^2_{\rm max} = &&(\sqrt s - m)^2\,,\\
Q^2_{{\scriptstyle\rm min \atop\scriptstyle\rm  max}} = &-&2 m^2 +  {1\over 2s}  
(s+m^2+P^2) (s+m^2-W^2)\nn\\
&\mp&
{1\over 2s}  
\Lambda(s,m^2,p^2) \Lambda(s,m^2,W^2)
\,.
\eeqa

At this stage we are ready to discuss the `equivalent bosons'
approximation for $s > P^2 \gg M^2_A, M^2_B, m^2$.
To this end we first observe that for given $P^2$ the hadronic part
$h_{\rho\sigma}(x,\tau)$ is logarithmically dominated by low $Q^2$ values:
\beqa
h_{\pm\mp}(x,\tau)&\simeq& x^2 \log{P^2 \over Q^2},\\
h_{\pm\pm}(x,\tau)&\simeq& (1-x)^2 \log{P^2 \over Q^2}
\eeqa
with other components finite for $Q^2/P^2 \rightarrow 0$.
Thus we will be interested in the region $Q^2 \ll P^2$ corresponding to
the leading logarithmic approximation.
In this limit
$x$ gains the 
meaning of the Bjorken variable of the hadronic subprocess:
\beq
x = {P^2\over 2pq}\,,
\eeq
while $y$ becomes the electroweak boson momentum fraction with respect
to~$l$:
\beq
y = {pq \over pl} = {W^2+P^2 \over s +P^2}\,.
\eeq
The product of $x$ and $y$ is independent of the integration variables
and reads
\beq
z \equiv xy = {P^2\over 2pl}\,.
\eeq

As shown in Ref.\cite{WS1}
the leading terms of $h_{\rho\sigma}$ are in fact the 
products of the logarithmic factors and the splitting functions
$P^\rho_{q_\eta }(x)$ and $P^\rho_{\bar q_{\eta}}(x)$
of a spin-1 boson of helicity $\rho$ into a quark (antiquark) of helicity
$\eta$: 
\beq
h_{\rho\sigma}(x,Q^2) = {1\over 6} 
[P^\rho_{q_{-\sigma}}(x) +  P^\rho_{\bar q_{-\sigma}}(x)]
\log{P^2 \over Q^2}
\eeq
with
\beqal
P^\pm_{q_\pm }(x) &=&  P^\pm_{\bar q_{\pm}}(x) = 3x^2\,,\\
P^\mp_{q_\pm }(x) &=&  P^\mp_{\bar q_{\pm}}(x) = 3 (1-x)^2\,.
\eeqal

Changing the integration variable $W^2$ to $y$ 
we obtain the formula for ${\J}_{\sigma}^{\lambda\eta}$ 
which allows for the discussion of the equivalent boson approximation:
\beqa
{\J}_{\sigma}^{\lambda\eta}\!\!&=&\!\!
 {\alpha_{\rm s} \alpha \over 6}
\sum_{\scriptstyle A,B=\atop \scriptstyle \gamma, \rm W, Z}
\;\sum_{\rho=\pm}
\int_z^{y_{\rm max}} {dy\over y} 
 {\alpha \over 2\pi}
  P^{\ell_\lambda}_{A_\rho B_\rho}(y)
\,
\hat g^{Aq}_\eta \hat g^{Bq}_\eta
\left[P^\rho_{q_{-\sigma}}\left({z\over y}\right) 
+  P^\rho_{\bar q_{-\sigma}}\left({z\over y}\right)\right]
\nn\\
&\times&
\int_{Q^2_{\rm min}}^{Q^2_{\rm max}} {Q^2 dQ^2 \over (Q^2 + M_A^2)(Q^2 + M_B^2)}
\,
\log{P^2 \over Q^2}
\,,
\label{J1}
\eeqa
where
\beqa
y_{\rm max} &=& 1 - {\cal O}(m^2/P^2)
\,,\\
Q^2_{\rm min} &=& m^2 {y^2 \over 1-y}
\,,\\
Q^2_{\rm max} &=& P^2 {z+y-zy \over z}
\,.
\eeqa
$P^{\ell_\lambda}_{A_\rho B_\rho}$ in the above equation is a generalization of the
electron-boson splitting function, which takes into account the
interference effects.

Let us note that the minimal value of $Q^2$ is the only place where the
electron mass must be kept finite to regularize collinear divergencies
in the case of photon exchange.
However, in order to remain within the leading logarithmic approximation we have to pay
particular attention to the upper limit of the $Q^2$ integration.
In general it is a function of $P^2$, however
integration up to the maximum kinematically allowed value $Q^2_{\rm max}$
would violate the condition $Q^2/P^2 \ll 1$. For our approximation
to work we must integrate over $Q^2$ up to some 
$\hat Q^2_{\rm max} \ll P^2$.
Integrating \Eq{J1} over $Q^2$ within such limits and keeping only 
leading-logarithmic terms
leads to
\def\Qbar{\bar{Q}}
\def\Qhsq{\hat Q^2_{\rm max}}
\beq
{\J}_{\sigma}^{\lambda\eta}
 =
 {\alpha_{\rm s} \alpha \over 6}
 \sum_{A,B,\rho} 
\hat  g^{Aq}_\eta  \hat g^{Bq}_\eta
 F^{\ell_\lambda}_{A_\rho B_\rho}(\Qhsq)
\otimes
[ P^\rho_{q_\eta } \delta_{\eta,-\sigma} +  P^\rho_{\bar q_{-\eta}} \delta_{\eta,\sigma}]
\log 
{P^2\over \Qbar^2}
\,.
\label{JLL}
\eeq
where $F^{\ell_\lambda}_{A_\rho B_\rho}(y,\Qhsq)$ is the density matrix of
polarized bosons inside the lepton and $\Qbar^2$ is an average $Q^2$ value.
(Exact result from \Eq{J1} is 
$\Qbar^2 = \sqrt{\hat Q^2_{\rm max} Q^2_{\rm min}}$ 
but within the leading-log approximation only the order of magnitude is relevant).

The transverse components of $F^{{\rm e}^-}_{A_\rho B_\rho}$ 
in the case of unpolarized electron
read (formulae for polarized electron are given in the Appendix B)
\begin{eqlettarray}
\eqalabel{bos}
  F^{\rm e^-}_{\gamma_\pm \gamma_\pm}(y,Q^2) &=&
  {\alpha \over 2\pi}
  {(1-y)^2 + 1 \over 2y}
  \;
  \log {Q^2\over m_e^2}
\label{bos1}
\,,\\
  F^{\rm e^-}_{{\rm Z}_+ {\rm Z}_+}(y,Q^2) &=&
  {\alpha \over 2\pi}
  \tthWsq \; {\rho_{\rm W}^2 (1-y)^2 + 1 \over 2y}
  \;
  \log{Q^2 +M_{\rm Z}^2\over M_{\rm Z}^2}
\,,\;\;\;\;\;\\
  F^{\rm e^-}_{{\rm Z}_- {\rm Z}_-}(y,Q^2) &=&
  {\alpha \over 2\pi}
  \tthWsq \; {\rho_{\rm W}^2 + (1-y)^2   \over 2y}
  \;
  \log{Q^2 +M_{\rm Z}^2\over M_{\rm Z}^2}
\,,\\
  F^{\rm e^-}_{\gamma_+ {\rm Z}_+}(y,Q^2) &=&
  {\alpha \over 2\pi}
  \tthW \; {\rho_{\rm W} (1-y)^2 - 1 \over 2y}
  \;
  \log{Q^2 +M_{\rm Z}^2\over M_{\rm Z}^2}
\,,\\
  F^{\rm e^-}_{\gamma_- {\rm Z}_-}(y,Q^2) &=&
  {\alpha \over 2\pi}
  \tthW \; {\rho_{\rm W} - (1-y)^2   \over 2y}
  \;
  \log{Q^2 +M_{\rm Z}^2\over M_{\rm Z}^2}
\,,\\
  F^{\rm e^-}_{{\rm W}_+ {\rm W}_+}(y,Q^2) &=&
  {\alpha \over 2\pi}
  {1\over 4\sthWsq} \; {(1-y)^2 \over y}
  \;
  \log{Q^2 +M_{\rm W}^2\over M_{\rm W}^2}
\,,\\
  F^{\rm e^-}_{{\rm W}_- {\rm W}_-}(y,Q^2) &=&
  {\alpha \over 2\pi}
  {1\over 4\sthWsq} \; {1 \over y}
  \;
  \log{Q^2 +M_{\rm W}^2\over M_{\rm W}^2}
\,,
\label{bos7}
\end{eqlettarray}
where
$\theta_W$ is the Weinberg angle 
and
\beq
  \rho_{\rm W} =  {1\over 2\sthWsq} -1 \,.
\eeq
\kom{
and
\beq
\log\mu_0 = \log {{\epsilon P^2}\over m_e^2}, \;
\log\mu_B = \log{{\epsilon P^2 +M_B^2}\over M_B^2}.
\label{logs}
\eeq
}
All other density matrix elements (containing at least one  longitudinal
boson) do not contribute at the leading-logarithmic level.

The density matrix elements $F^{\rm e^-}_{A_\rho B_\rho}(y,Q^2)$
should be considered as a generalized solution to the \Eq{meq-B}.
At this point we are able to define the splitting functions of an
electron into a quark at the momentum scale $Q^2$ as
\beq
  P^{\rm e^-}_{q_\eta}(Q^2) =
  \sum_{AB}  \hat g^{Aq}_\eta  \hat g^{Bq}_\eta
  \sum_\rho F^{\rm e^-}_{A_\rho B_\rho} (Q^2)\otimes P^\rho_{q_\eta}\,.
\label{sf-def}
\eeq
This is the generalization of the first sum of \Eq{meq-h}.
The explicit expressions for quarks read (see Appendix B for the
polarized electron case)
\beqal
  P^{\rm e^-}_{q_+}(z,Q^2) &&=  {3\alpha \over 4\pi} \left\{
  e_q^2  [ \Phi_+(z) +  \Phi_-(z) ]
  \log {Q^2\over m_e^2}\right.
\nn\\
  + e_q^2 &&\tan^4\theta_{\rm W} [ \Phi_+(z) + \rW^2 \Phi_-(z) ]
  \log{Q^2 +M_{\rm Z}^2\over M_{\rm Z}^2}
\nn\\
  - 2e_q^2 &&
\left.
\tan^2\theta_{\rm W} [ -\Phi_+(z) + \rW \Phi_-(z) ]  
  \log{Q^2 +M_{\rm Z}^2\over M_{\rm Z}^2}
\right\} \,,\\
  P^{\rm e^-}_{q_-}(z,Q^2) &&=  {3\alpha \over 4\pi} \left\{
  e_q^2  [ \Phi_+(z) +  \Phi_-(z) ]  
  \log {Q^2\over m_e^2}\right.
\nn\\
  + z_q^2&& \tan^4\theta_{\rm W} [ \Phi_-(z) + \rW^2 \Phi_+(z) ]  
  \log{Q^2 +M_{\rm Z}^2\over M_{\rm Z}^2}
\nn\\
  + 2e_q z_q&& \tan^2\theta_{\rm W} [ -\Phi_-(z) + \rW \Phi_+(z) ]  
  \log{Q^2 +M_{\rm Z}^2\over M_{\rm Z}^2}
\nn\\
  +&&
\left.
 (1+\rW)^2 \Phi_+(z) \delta_{q \rm d}  
  \log{Q^2 +M_{\rm W}^2\over M_{\rm W}^2}
\right\} \,,
\eqalabel{spf}
\eeqal
where
\beqa
\Phi_+(z) &=& {1-z \over 3 z} (2+ 11z +2z^2) + 2(1+z)\log z\,,\\
\Phi_-(z) &=& {2 (1-z)^3 \over 3 z}\,,
\label{Phi-def}
\eeqa
and
\beq
z_q = {T_3^q \over \sthWsq} - e_q \,,
\label{zq-def}
\eeq
with $e_q$ and $T_3^q$ being the quark charge and 3-rd weak isospin component, 
respectively.
The splitting functions for an antiquark of the opposite helicity can be
obtained from \Eq{spf} by interchanging $\Phi_+$ with $\Phi_-$.

These new splitting functions depend on the scale $\Qhsq$. This scale can be
in principle fixed by experimental cuts, in such case the final lepton must
be tagged.
A condition commonly used in phenomenological applications
\cite{Fritj} is
$\hat Q^2_{\rm max} = \epsilon P^2$,  where $\epsilon \ll 1$
and in general depends on $y$ and $z$%
In this case the electron splitting functions depend on 
$\log P^2$ and hence on $t$.
In the following we will concentrate on 
the case with the maximum virtuality being $P^2$-dependent.
In this case the master equations \Eq{meq-h} become
\beqal
\eqalabel{meq2}
{d\over dt}F^{{\rm e}^-_\lambda}_{q_\eta/\bar q_\eta}(t) &=&
{\alpha\over 2\pi} 
P^{{\rm e}^-_\lambda}_{q_\eta/\bar q_\eta}(t)
+
{\alfs(t)\over 2\pi} 
\sum_{\scriptstyle h'=\atop \scriptstyle {\rm q, \bar q, G}} 
\sum_{\rho=\pm 1}
P_{q_\eta/\bar q_\eta}^{h'_\rho}
 \otimes
 F_{h'_\rho}^{{\rm e}^-_\lambda}(t)
\label{meq-q}
\,,\;\;\;\;\;
\\
{d\over dt}F^{{\rm e}^-_\lambda}_{G_\eta}(t) &=&
{\alfs(t)\over 2\pi} 
\sum_{\scriptstyle h'=\atop \scriptstyle {\rm q, \bar q, G}} 
\sum_{\rho=\pm 1}
P_{G_\eta}^{h'_\rho}
 \otimes
 F_{h'_\rho}^{{\rm e}^-_\lambda}(t)
\label{meq-G}
\,.
\eeqal

Note that the convolution of the equivalent boson distributions and
boson-quark
splitting functions, \Eq{sf-def}, occurs at the level of splitting 
functions. It is not equivalent to the usually performed convolution of
the distribution functions because of the $P^2$ dependence of the boson
densities \Eq{bos}. 
Only in the case when the upper limit
of integration
$\Qhsq$ is kept fixed ($P^2$-independent), 
are the convolutions equivalent at both levels.

To summarize, our splitting functions show two new features. The first one, 
already mentioned before, is
the contribution from the interference of  electroweak bosons ($\gamma$ and 
$Z$ only). 
The second one is their $P^2$ dependence, which results from
the upper integration limit $\hat Q^2_{\rm max}$. 
\kom{
Having completed the first step of our procedure --- the calculation of the 
splitting functions, we can proceed to the resumation of the QCD cascade
using the evolution equations.
}

\section{Asymptotic solutions to the master equations}
The master equations \Eq{meq2}, with spin dependent functions,  
are most conveniently solved in terms of polarized and
unpolarized distributions \cite{WS1}:
\beqal
F_{h}^{{\rm e}^-_\lambda} &=&
 F_{h_+}^{{\rm e}^-_\lambda} + F_{h_-}^{{\rm e}^-_\lambda}
\,,\\
\Delta F_{h}^{{\rm e}^-_\lambda} &=&
 F_{h_+}^{{\rm e}^-_\lambda} - F_{h_-}^{{\rm e}^-_\lambda}
\,,
\eeqal
\beqal
P_{h}^{{\rm e}^-_\lambda} &=&
 P_{h_+}^{{\rm e}^-_\lambda} + P_{h_-}^{{\rm e}^-_\lambda}
\,,\\
\Delta P_{h}^{{\rm e}^-_\lambda} &=&
 P_{h_+}^{{\rm e}^-_\lambda} - P_{h_-}^{{\rm e}^-_\lambda}
\eeqal
and
\beqal
P_{h}^{h'} &=& P_{h_+}^{h'_+} + P_{h_-}^{h'_+}
\,,\\
\Delta P_{h}^{h'} &=& P_{h_+}^{h'_+} - P_{h_-}^{h'_+}
\,.
\eeqal
Upon substitution of the above definitions into \Eq{meq2} we obtain two
sets of
master equations where unpolarized and polarized functions do not mix:
\beqal
\eqalabel{meq-pol}
{d\over dt}F^{{\rm e}^-_\lambda}_{h}(t) &=&
{\alpha\over 2\pi} 
P^{{\rm e}^-_\lambda}_{h}(t)
+
{\alfs(t)\over 2\pi} 
\sum_{\scriptstyle h'=\atop \scriptstyle {\rm q, \bar q, G}} 
P_{h}^{h'} \otimes F_{h'}^{{\rm e}^-_\lambda}(t)
\,,\\
{d\over dt}\Delta F^{{\rm e}^-_\lambda}_{h}(t) &=&
{\alpha\over 2\pi} 
\Delta P^{{\rm e}^-_\lambda}_{h}(t)
+
{\alfs(t)\over 2\pi} 
\sum_{\scriptstyle h'=\atop \scriptstyle {\rm q, \bar q, G}} 
\Delta P_{h}^{h'} \otimes \Delta F_{h'}^{{\rm e}^-_\lambda}(t)
\,,
\eeqal
In the following we will present the asymptotic solutions to these
equations for the case of unpolarized electron, 
$F^{{\rm e}^-}_h$ and $\Delta F^{{\rm e}^-}_h$.

The equations \Eq{meq2} have been derived within the leading
logarithmic approximation which is justified for 
asymtotically large $P^2$ values. In this region all the logarithms of $P^2$
over any finite scale are approximately equal and 
in the QCD evolution we choose them to be scaled by $\Lambda_{\rm QCD}$:
\beq
t=\log{P^2\over \Lambda_{\rm QCD}^2}\,.
\eeq
We present below the solution to the master equations in this asymptotic
(large $t$) region.

%
\begin{figure}[h]
\begin{center}
\mbox{\vbox{%
\def\epsfsize#1#2{\figscale#1}
\epsfbox{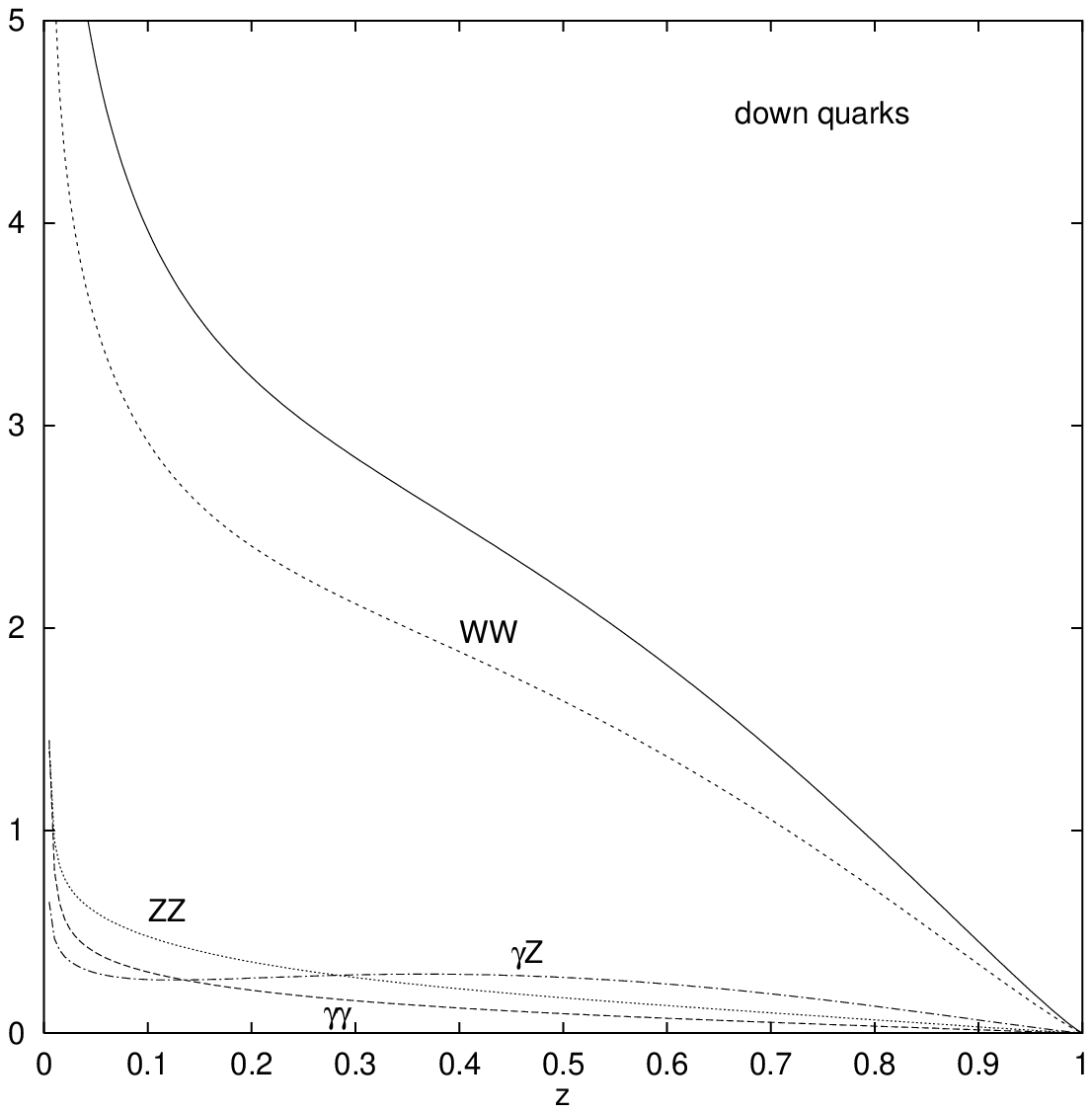}
\epsfbox{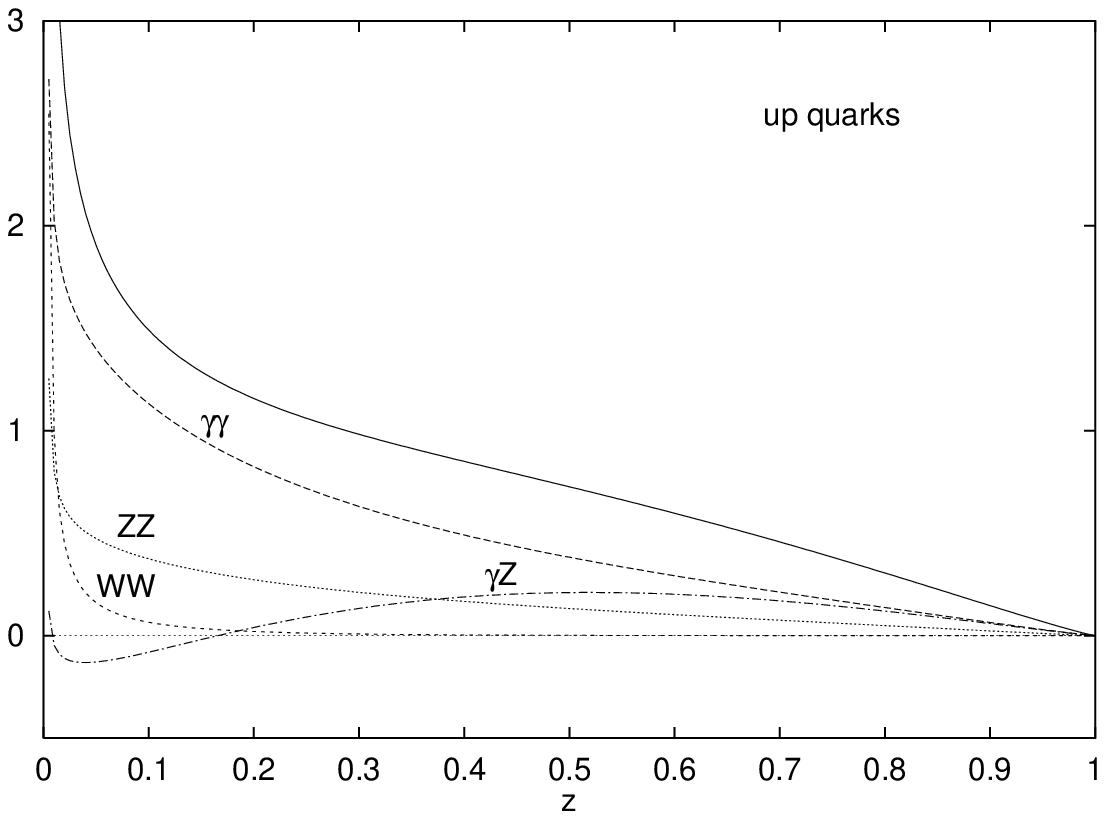}
}}
\end{center}
\caption{Unpolarized quark distributions $z f^{\rm as}_{\rm u/d}(z)$ --- solid line.
The other lines show contributions from different electroweak bosons.}
\label{f-qdistr}
\end{figure}
%
%
\begin{figure}[h]
\begin{center}
\mbox{\vbox{%
\def\epsfsize#1#2{\figscale#1}
\epsfbox{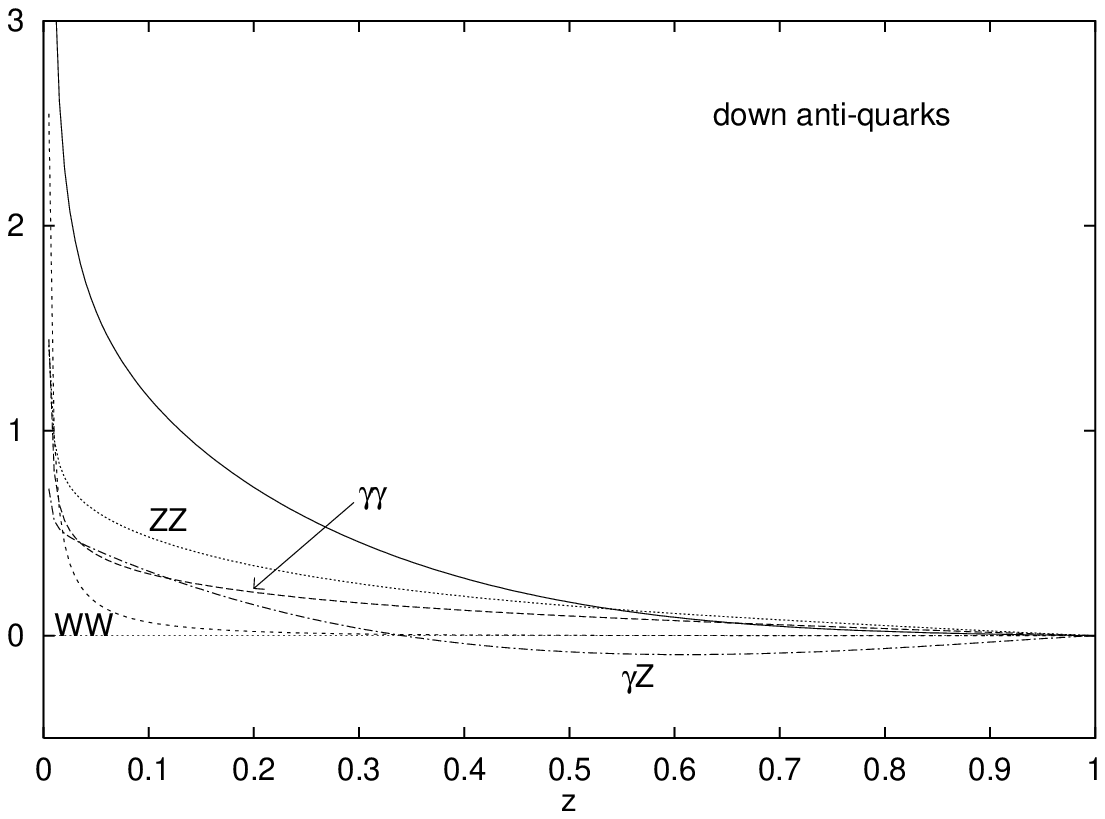}
\epsfbox[67 47 411 395]{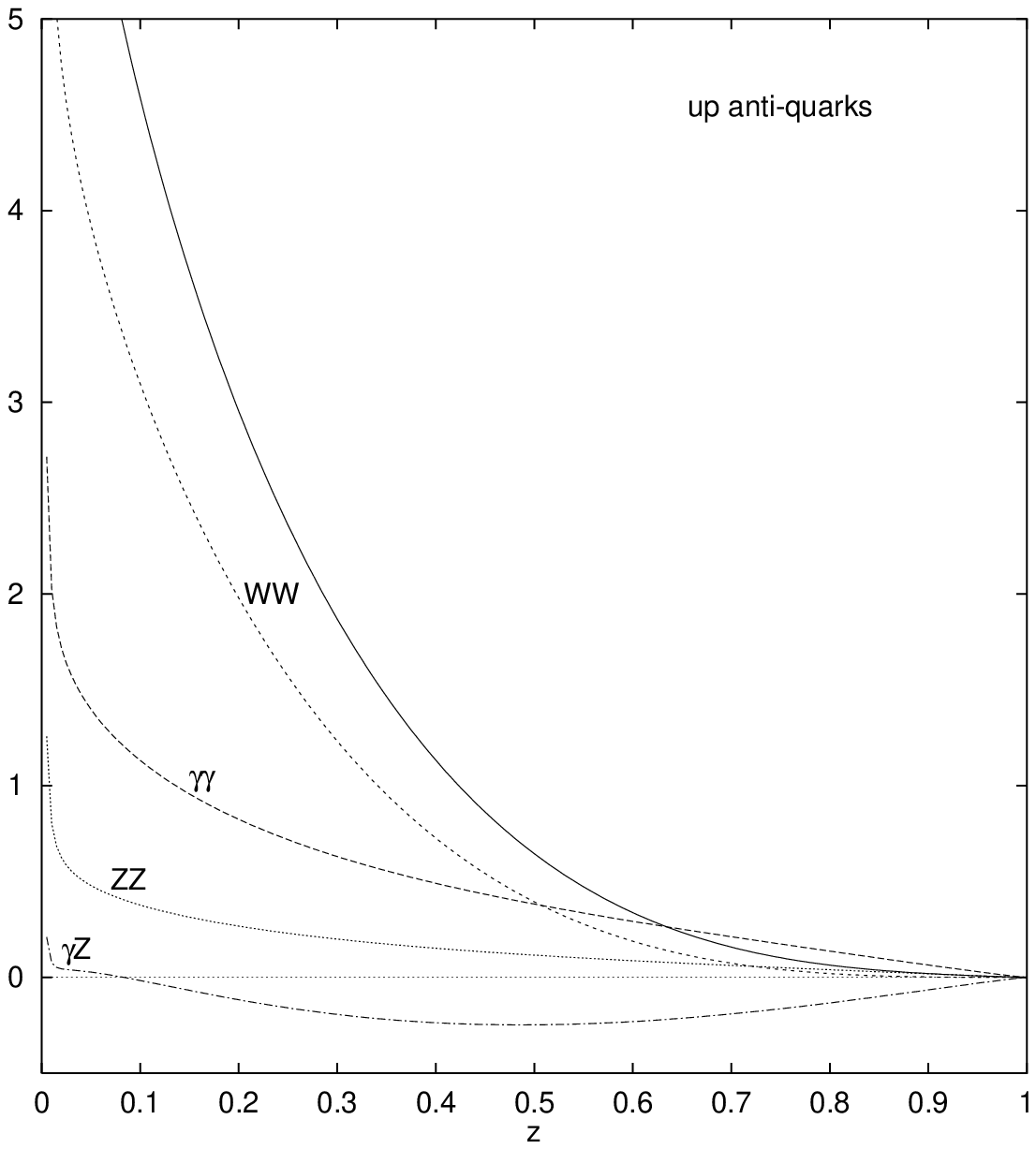}
}}
\end{center}
\caption{Unpolarized anti-quark distributions 
$z f^{\rm as}_{\rm \bar u/\bar d}(z)$ --- solid line.
The other lines show contributions from different electroweak bosons.}
\label{f-aqdistr}
\end{figure}
%
%
\begin{figure}[h]
\begin{center}
\mbox{\vbox{%
\def\epsfsize#1#2{0.73#1}
\epsfbox[50 60 347 280]{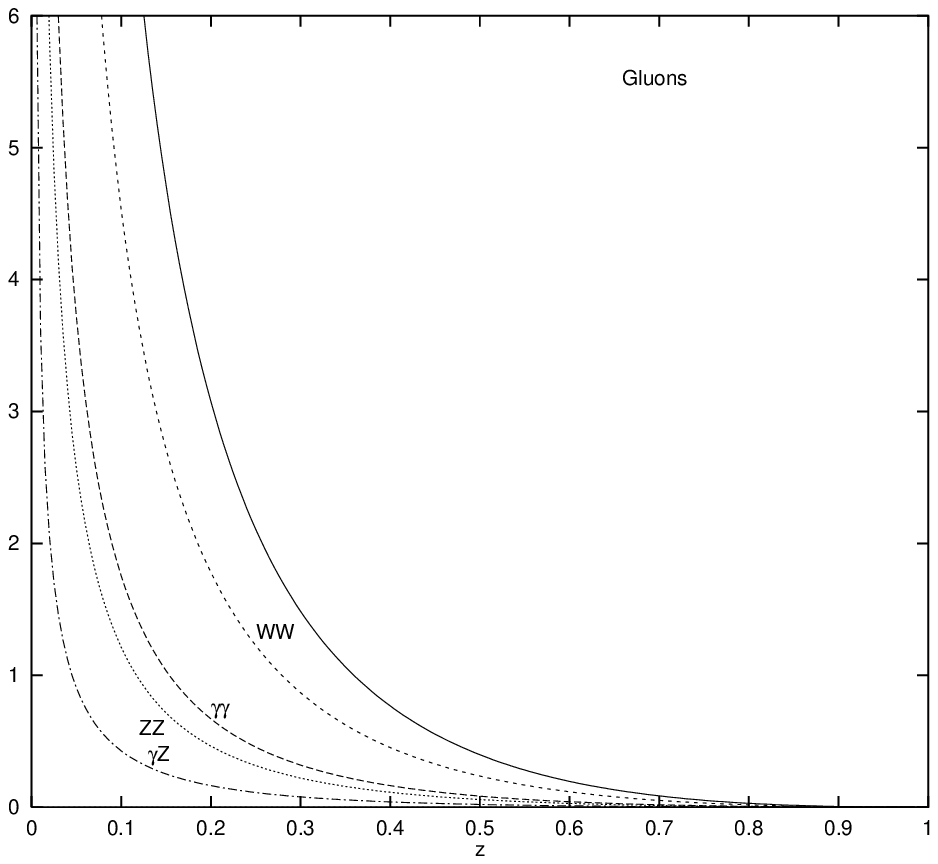}%
}}
\end{center}
\caption{Unpolarized gluon distribution $z f^{\rm as}_{\rm G}(z)$ --- solid line.
The other lines show contributions from different electroweak bosons.}
\label{f-Gdistr}
\end{figure}
%
%
\begin{figure}[h]
\begin{center}
\mbox{\vbox{%
\def\epsfsize#1#2{\figscale#1}
\epsfbox[50 60 410 365]{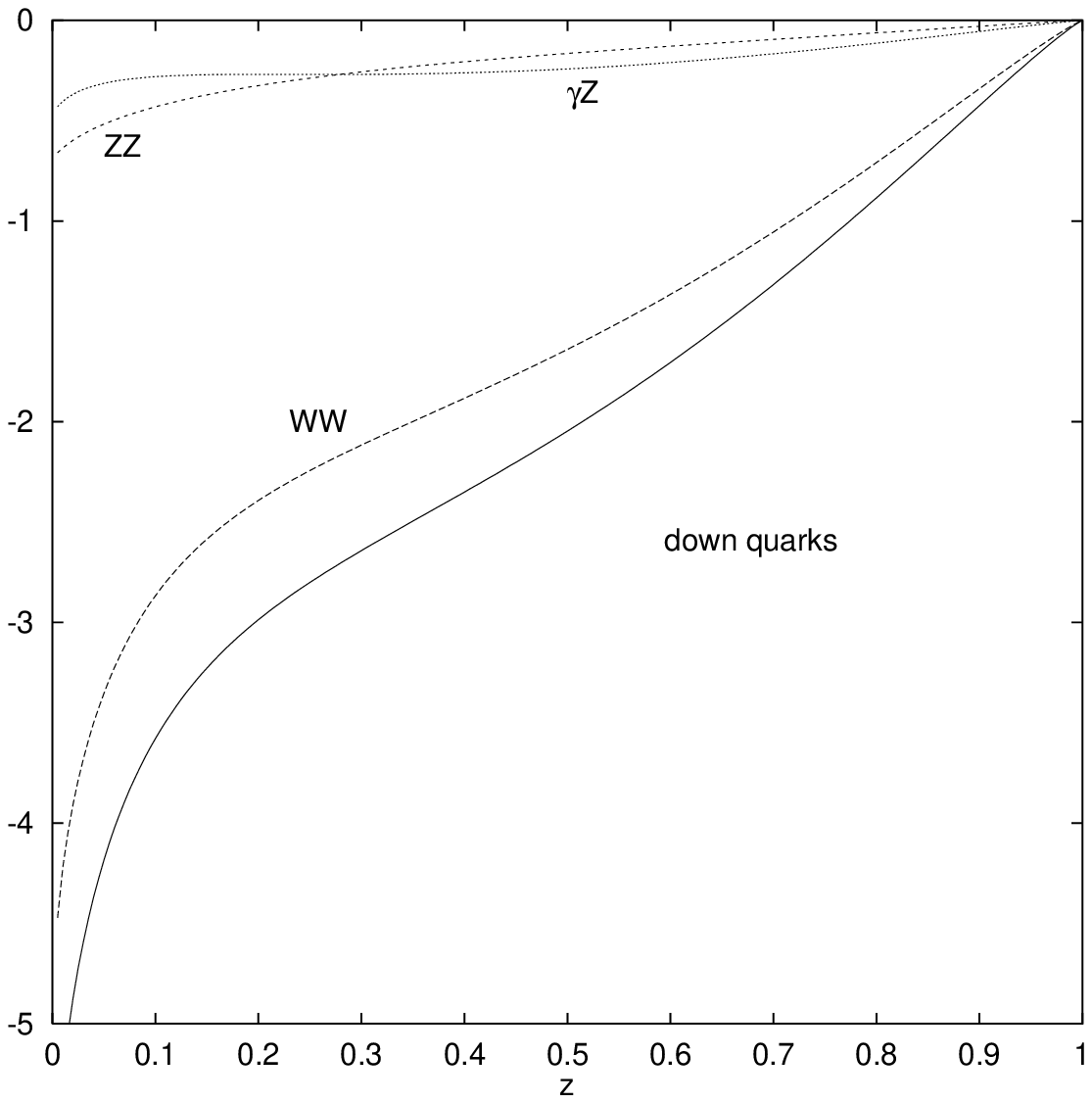}
}}
\end{center}
\caption{Polarized d-quark distribution $z\Delta f^{\rm as}_{\rm d}(z)$ --- solid line.
The other lines show contributions from different electroweak bosons.}
\label{f-dqdistr}
\end{figure}
%

%
\begin{figure}[h]
\begin{center}
\mbox{\vbox{%
\def\epsfsize#1#2{0.78#1}
\epsfbox[50 60 347 280]{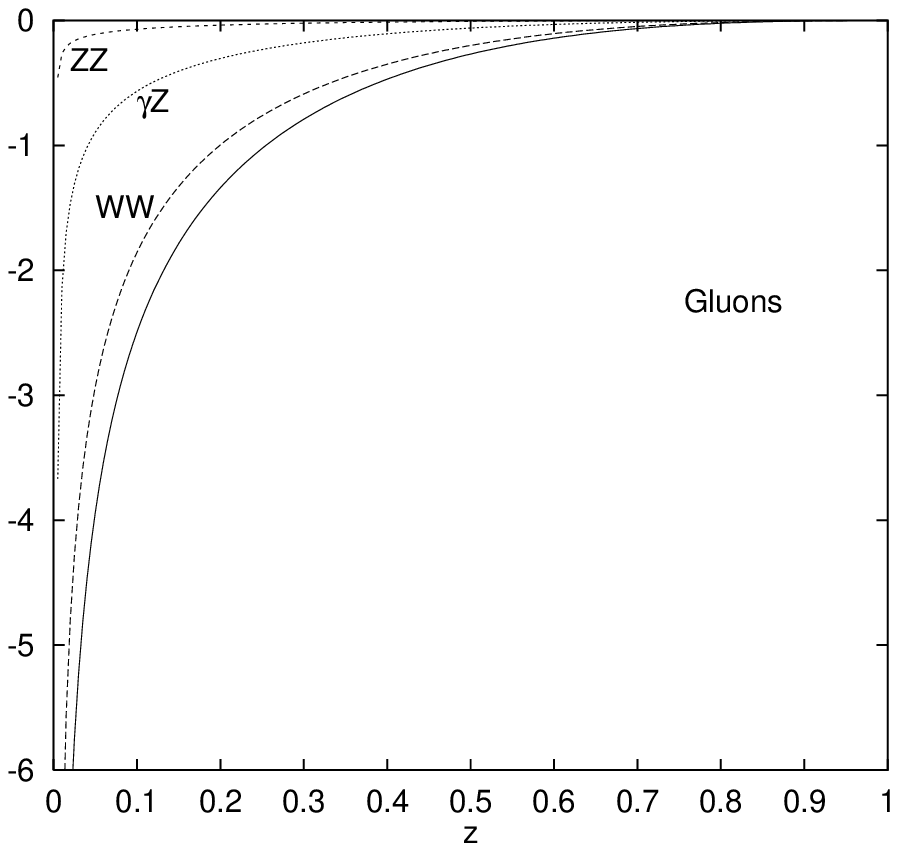}%
}}
\end{center}
\caption{Polarized gluon distribution $z\Delta f^{\rm as}_{\rm G}(z)$ --- solid line.
The other lines show contributions from different electroweak bosons.}
\label{f-dGdistr}
\end{figure}

%

The strong coupling constant is approximated by
\begin{equation}
\alpha_{\rm s}(t) \simeq {2 \pi \over bt}\,,
\label{astr}
\end{equation}
with $b=11/2 - n_{\rm f}/3$ for $n_{\rm f}$ flavours.
Parametrizing the asymptotic solution to Eqs.(\ref{meq-pol}) for the
QCD parton $h$ in the unpolarized electron as
\beqal
F_{h}^{{\rm e}^-}(z,t) &\simeq& {1\over 2} 
\left( {\alpha \over 2\pi} \right)^2 
f^{\rm as}_{h}(z) \, t^2
\,,\\
\Delta F_{h}^{{\rm e}^-}(z,t) &\simeq& {1\over 2} 
\left( {\alpha \over 2\pi} \right)^2 
\Delta f^{\rm as}_{h}(z) \, t^2
\eqalabel{asdef}
\eeqal
we obtain purely integral equations
\beqal
f^{\rm as}_{h} &=& \hat P_{h}^{{\rm e}^-}
    + {1\over 2b} \sum_{h'} P_{h}^{h'}
 \otimes f^{\rm as}_{h'}
\,,\\
\Delta f^{\rm as}_{h} &=& \Delta \hat P_{h}^{{\rm e}^-}
    + {1\over 2b} \sum_{h'} \Delta P_{h}^{h'}
 \otimes \Delta f^{\rm as}_{h'} 
\,,
\eeqal
where $\hat P_{h}^{{\rm e}^-}(z)$ and $\Delta\hat P_{h}^{{\rm e}^-}(z)$ 
are equal to  $P_{h}^{{\rm e}^-}(z)$ and $\Delta P_{h}^{{\rm e}^-}(z)$,
respectively,
with all logs put to 1.

We solve the above equations by a numerical procedure described 
in Ref.\thinspace\cite{WS1}. We take number of flavours $n_{\rm f}=5$.
In the asymptotic region, where the quark masses can be neglected, we
are left with only two different quark distributions: up-type and
down-type.

In figures \ref{f-qdistr}, \ref{f-aqdistr} and \ref{f-Gdistr} 
we present the unpolarized quark, anti-quark and gluon distributions.
The polarized distributions are depicted in figures \ref{f-dqdistr} and
\ref{f-dGdistr}. The density $z\Delta f^{\rm as}_{\rm u}(z)$, not shown here,
 is
more than 10 times smaller than $z\Delta f^{\rm as}_{\rm d}(z)$.
One notices significant contribution from the  W
intermediate state in the density of d-type and $\bar{\rm u}$-type quarks.
The most surprising however is the $\gamma$-Z
interference contribution  which cannot be neglected, as it is comparable 
to the Z term. It violates the standard probabilistic approach
where only diagonal terms are taken into account.  This also stresses 
 the necessity  
of introducing the concept
of electron structure function in which all 
contributions from intermediate bosons are properly summed up.
Due to the nature of weak couplings they 
turn out to be nonzero, 
even in the case of gluon distributions. Again the $\gamma$-Z interference
term  is important and the W contribution dominates in the asymptotic region.  

One should keep in mind that at finite $t$ the logarithms multiplying the
photon contribution differ  from the remaining ones (cf. \Eq{bos1}). Being scaled
by $m_e$, they lead to the photon domination at presently available $P^2$. The 
importance of the interference term remains constant relative to the $Z$ contribution,
as they are both governed by the same logarithm. 


\section{Conclusions.}
In the paper we investigated the hadronic content of the electron defining
a new quantity --- the electron structure function. We defined it and 
constructed its evolution equations. We also gave asymptotic solutions
to these equations demonstrating rich flavour and spin dependence. 
In all cases the parton densities inside the electron grow as logarithm
squared of the external momentum scale (in our calculation the
gluon momentum $P^2$). This construction, which looks
at first sight a formal manipulation only, brings in several new issues.
The most important ones are the correct treatment of the intermediate
electroweak bosons and precise formulation of the structure function 
evolution. In the first case one faces the necessity of introducing 
the electroweak boson density matrix with the off-diagonal ($\gamma$-Z)
terms comparable to the diagonal ones. The probabilistic approach,
one is used to in the partonic picture, can be restored only when looking
at the whole process \ie an electron emitting quarks and gluons.
Also the QCD evolution can be treated correctly only when looking
at the complete system. The virtuality of the intermediate boson
depends in general on the external momentum  and in this way the
electroweak sector enters the QCD evolution. The derivation of the 
electron structure function presented above assumes a totally inclusive
situation when the outgoing lepton is not tagged. In the opposite situation,
which is possible only with neutral currents and away from the forward
direction, the virtuality of the intermediate 
boson can be controlled by experiment and kept independent of the external
momentum scale. In all cases the known boson structure function should
be treated with great care in the lepton induced processes. The question 
becomes very important at momenta where all electroweak bosons come into play.
\begin{figure}[t]
\begin{center}
\mbox{\vbox{%
\def\epsfsize#1#2{\figscale#1}
\epsfbox{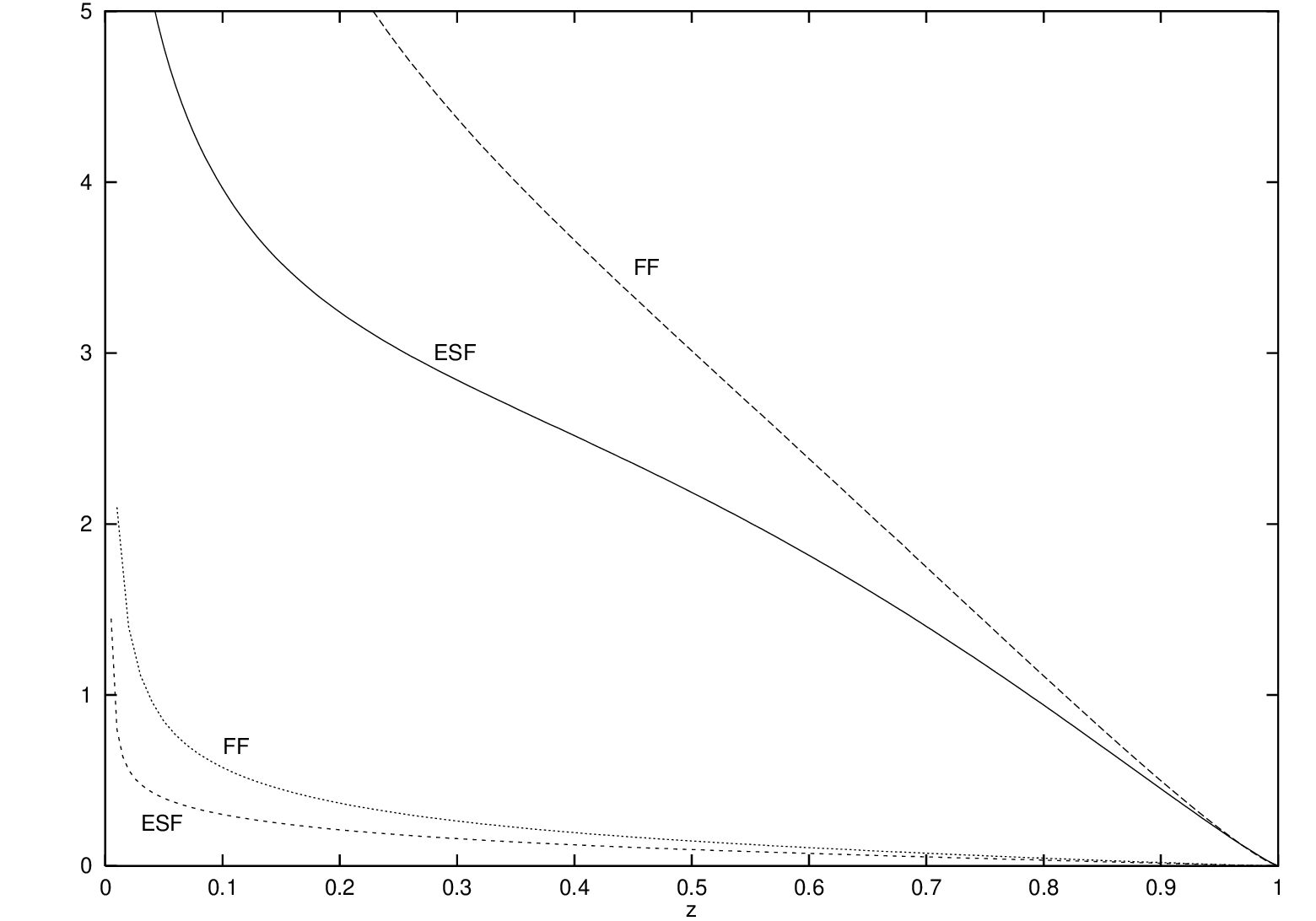}%
}}
\end{center}
\caption{Comparison
of unpolarized d-quark distributions $z f_{\rm d}^{\rm as}(z)$
calculated by ESF and FF methods. The upper two lines result from
contributions from all electroweak bosons while the lower
two from $\gamma\gamma$ only.}
\label{f-dscmp}
\end{figure}
But even at presently 
available momenta, where the 
`resolved' photon dominates, one can see how the correct treatment of the 
scales changes the shape of parton distributions. 
In Fig.\thinspace\ref{f-dscmp} we compare the asymptotic solutions
of the evolution equations 
following from our procedure (ESF) with those following from naive 
application of the convolution \Eq{FF} (FF). It is possible that 
the difference can be traced in the analysis of presently available data.
The inclusion of  effects connected with finite momenta and construction of 
the electron structure function which could be applied to present experiments
is a natural next step along the proposed lines.

\bigskip\noindent{\bf Acknowledgements}

This work has been partially performed during our visit to Brookhaven National
Laboratory and DESY. The authors would like to thank the
Theory Groups of BNL and DESY for their warm hospitality.

\section*{Appendix A}
\reseteqprefix{A.}
\def\LOG{{\cal L}}
\def\tovtb{\tilde\tau}
\def\xb{(1-x)}

Here we present full results for the hadronic part of the 
QCD current matrix element squared as well as some kinematical relations.

The $h_{\rho\sigma}(x,\tau)$ functions of \Eq{Hres} read:
\arraycolsep=2pt
\begin{equation}
\begin{array}{rcl}
  h_{--}(x,\tau) &=&  h_{++}(x,\tau) =  \xb^2 \LOG
\\
  &\displaystyle{ +  {\xb^2 \over 16 x^2}  }
  &\left\{  - \tau ( 1 + x )^2 +  \tovtb  ( 1 - 14 x + 49 x^2 )\right.
\\
  &&     + 3  \tovtb^2 ( 1 - 14 x + 25 x^2) 
\\
  && + 3  \tovtb^3 ( 3 - 18 x - 19 x^2 )
\\
   &&\left. + 3  \tovtb^4 \xb  ( 3 -7 x ) +3  \tovtb^5 \xb^2 \right\}\LOG       
\\
  &\displaystyle{ +{\xb^2 \over 8 x^2}}
  &\left\{ 8\tovtb x ( 2 - 5 x) - 4 \tovtb^2   (1 - 8 x + 10 x^2)\right.
\\
  &&\left. - 6 \tovtb^3  \xb (1 - 3 x )  - 3 \tovtb^4 \xb^2 \right\}
\\
  &&   - 2 (1 - x)^2
\,,
\end{array}
\end{equation}
\begin{equation}
\begin{array}{rcl}
     h_{-+}(x,\tau) &=&  h_{+-}(x,\tau) = x^2 \LOG
\\ 
  &\displaystyle{ +  {1 \over 16 x^2}  }
  &\left\{ -\tau   \xb^2  ( 1 + x)^2      \right.
\\
  && + \tovtb   ( 1 - 2 x^2 - 32 x^3 + 49 x^4 )
\\
  &&  + \tovtb^2  \xb   ( 3 - 13 x + 53 x^2 - 75 x^3 ) 
\\
  && + \tovtb^3  \xb^2   ( 9 - 38 x + 57 x^2 )
\\
  &&\left. + 3  \tovtb^4  \xb^3   ( 3 - 7 x )  + 3  \tovtb^5   \xb^4  \right\}\LOG
\\
  &\displaystyle{ +{1 \over 8 x^2}}
  &\left\{ - 8 \tovtb  \xb   x^2  ( 1 - 5 x )   \right.
\\
  && -4  \tovtb^2  \xb^2   ( 1 - 4 x + 10 x^2 )
\\
  &&\left.   -6  \tovtb^3  \xb  3   ( 1 - 3 x ) -3   \tovtb^4  \xb^4  - 16 x^4  \right\}
\,,
\end{array}
\end{equation}

\begin{equation}
\begin{array}{rcl}
  h_{00}(x,\tau)
  &\displaystyle{ =  {\xb^2 \over 4 x^2}  }
  &\left\{ - \tau    (1 + x)^2   + \tovtb   ( 1 + 2 x + 17 x^2 ) \right.
\\
  &&   - \tovtb^2   ( 1 + 18 x - 55 x^2 )  + \tovtb^3   ( 5 - 46 x + 53 x^2 )
\\
  &&\left.   + 3  \tovtb^4  \xb   ( 3 - 7 x ) + 3 \tovtb^5  \xb^2 \right\}\LOG
\\
  &\displaystyle{ +{3 \xb^2 \over 2 x^2}}
  &\left\{ -8 \tovtb   x^2 +4  \tovtb^2  x  ( 2 - 3 x ) \right.
\\
   &&\left.    -2  \tovtb^3  \xb  ( 1 - 3 x ) - \tovtb^4  \xb^2 )\right\}
\,,
\end{array}
\end{equation}

\begin{equation}
\begin{array}{rcl}
  h_{0-}(x,\tau) &=&  h_{0+}(x,\tau) =  h_{-0}(x,\tau) =  h_{+0}(x,\tau)
\\
  &\displaystyle{ =  {1-x \over 8 x^2}  }
  &\left\{ \tau   \xb  (1+x)^2 - \tovtb   ( 1 + x - 17 x^2 + 31 x^3 ) \right.
\\
 && + \tovtb^2  ( 1 - 23 x + 79 x^2 - 65 x^3 ) 
\\
 && + \tovtb^3  \xb  ( 7 - 46 x + 55 x^2 )
\\
 &&\left. + 3 \tovtb^4  \xb^2   ( 3 - 7 x ) + 3 \tovtb^5   \xb^3 \right\} \LOG
\\
  &\displaystyle{ +{ 1-x \over 4 x^2}}
  &\left\{ + 4 \tovtb  x  ( 1 - 7 x + 8 x^2 )  \right.
\\
  && -2  \tovtb^2  \xb   ( 1 - 12 x + 19 x^2 )
\\
 &&\left. -6  \tovtb^3  \xb^2   ( 1 - 3 x ) -3  \tovtb^4   \xb^3 + 8 x^3  \right\}
\,,
\end{array}
\end{equation}
where
\beqa
\LOG &=& \log{P^2\over x^2 Q^2}\,,
\\
\tovtb &=& {\tau \over 1 -\tau}
\eeqa
and 
$\tau$ and $x$ are defined by \Eq{tau} and \Eq{xdef}, respectively.

For fixed $Q^2$ $h_{\rho\sigma}(x,\tau)$ are functions of $x$ only with
\beq
\tau = {x^2 Q^2 \over P^2 + x^2 Q^2}\,.
\eeq

Introducing $\bar s \equiv s + P^2 -m^2 =2lp$ we can define dimensionless variables
\beqa
\xi &=& {Q^2 \over \bar s}\,,
\\
z &=& {P^2 \over \bar s}
\eeqa
in terms of which we can write the formulae needed to change $(x,\tau)$
variables to $(y,\xi)$ used in the integral (\ref{J1}):
\beqa
\tau &=& {x^2 \xi \over z + x^2 \xi}\,,
\\
x &=& {\sqrt{y^2 +4(1-y) z \xi} - y \over 2(1-y)\xi}
\,.
\eeqa

\section*{Appendix B}
\reseteqprefix{B.}
In this appendix we present the formulae for the boson density matrix
elements and electron to quark splitting functions for the fully
polarized case.
As can be seen from \Eq{J1} and \Eq{JLL} the boson density matrix elements,
$F^{{\rm e}^-_\lambda}_{A_\rho B_\rho}$ are equal to 
$P^{{\rm e}^-_\lambda}_{A_\rho B_\rho}$ times appropriate logarithms.
The transverse components of
$P^{{\rm e}^-_\lambda}_{A_\rho B_\rho}$
can be read off \Eq{Pdef} and \Eq{Lres} and are equal to
\beq
  P^{{\rm e}^-_\pm}_{A_\rho B_\rho} =
  \hat g^{A{\rm e}}_\pm \hat g^{B{\rm e}}_\pm\,
  yY_{\pm\rho}(y)
  \,.
\eeq
\kom{
The couplings $\hat g^{A{\rm e}}_\pm$ are summarized in the following
table:
\begin{center}
\begin{tabular}[t]{@{~\vbox{\kern12pt}}lcc}
$AB$ & $\hat g^{A{\rm e}}_+$ & $\hat g^{B{\rm e}}_-$ \\
\hline
$\gamma\gamma$ & 1 & 1 \\
ZZ & $\tthWsq $ & $\rho_{\rm W}^2 \tthWsq $ \\
$\gamma$Z & $-\tthW $ & $\rho_{\rm W}\tthW $\\
WW & 0 & $\displaystyle{1+\rho_{\rm W} =  {1\over 2\sthWsq}} $
\end{tabular}
\end{center}
\medskip

Using these couplings we derive following formulae for
}
Inserting electroweak couplings we derive following formulae for
the transverse components of $F^{{\rm e}^-_\lambda}_{A_\rho B_\rho}$ 
in the case of polarized electron:
\Here
\begin{eqlettarray}
\eqalabel{bospol}
  F^{{\rm e}^-_\pm}_{\gamma_\rho \gamma_\rho}(y,Q^2) &=&
  {\alpha \over 2\pi}\, yY_{\pm\rho}(y)
  \,
  \log {Q^2\over m_e^2}
\,,\\
\kom{
  F^{{\rm e}^-_+}_{\gamma_\rho \gamma_\rho}(y,Q^2) &=&
  {\alpha \over 2\pi}\, yY_{\rho}(y)
  \,
  \log {Q^2\over m_e^2}
\,,\\
}
  F^{{\rm e}^-_-}_{{\rm Z}_\rho {\rm Z}_\rho}(y,Q^2) &=&
  {\alpha \over 2\pi}
  \rho_{\rm W}^2 \tthWsq 
  \, yY_{-\rho}(y)
  \,
  \log{Q^2 +M_{\rm Z}^2\over M_{\rm Z}^2}
\,,\\
  F^{{\rm e}^-_+}_{{\rm Z}_\rho {\rm Z}_\rho}(y,Q^2) &=&
  {\alpha \over 2\pi}
  \tthWsq 
  \, yY_{\rho}(y)
  \,
  \log{Q^2 +M_{\rm Z}^2\over M_{\rm Z}^2}
\,,\\
  F^{{\rm e}^-_-}_{\gamma_\rho {\rm Z}_\rho}(y,Q^2) &=&
  {\alpha \over 2\pi}
  \rho_{\rm W} \tthW
  \, yY_{-\rho}(y)
  \,
  \log{Q^2 +M_{\rm Z}^2\over M_{\rm Z}^2}
\,,\\
  F^{{\rm e}^-_+}_{\gamma_\rho {\rm Z}_\rho}(y,Q^2) &=&
  -{\alpha \over 2\pi}
  \tthW 
  \, yY_{\rho}(y)
  \,
  \log{Q^2 +M_{\rm Z}^2\over M_{\rm Z}^2}
\,,\\
  F^{{\rm e}^-_-}_{{\rm W}_\rho {\rm W}_\rho}(y,Q^2) &=&
  {\alpha \over 2\pi}
  {1\over 2\sthWsq} 
  \, yY_{-\rho}(y)
  \,
  \log{Q^2 +M_{\rm W}^2\over M_{\rm W}^2}
\,,\\
  F^{{\rm e}^-_+}_{{\rm W}_\rho {\rm W}_\rho}(y,Q^2) &=& 0
\,,
\end{eqlettarray}
where
\beq
 yY_-(y) = {(1-y)^2\over y}
\,,\;\;\;\;\;
 yY_+(y) = {1\over y}
\,.
\eeq

The splitting functions of a polarized electron into  
a polarized quark are defined by \Eq{sf-def} with 
$F^{{\rm e}^-_\lambda}_{A_\rho B_\rho}$ taken from \Eq{bospol} and a
straightforward calculation leads to the following result:
\Here
\beqal
\eqalabel{spfpol}
  P^{{\rm e}^-_-}_{q_+}(z,Q^2) &=&  {3\alpha \over 2\pi} \Phi_-(z) \left\{
  e_q^2
  \log {Q^2\over m_e^2}
  \right.
\nn\\
  &+& e_q^2 \rW^2 \tan^4\theta_{\rm W}
  \log{Q^2 +M_{\rm Z}^2\over M_{\rm Z}^2}
\nn\\
  &-& \left. 2e_q^2
  \rW \tan^2\theta_{\rm W}
  \log{Q^2 +M_{\rm Z}^2\over M_{\rm Z}^2}
\right\} \,,\\
%
  P^{{\rm e}^-_+}_{q_+}(z,Q^2) &=&  {3\alpha \over 2\pi} \Phi_+(z) \left\{
  e_q^2
  \log {Q^2\over m_e^2}
  \right.
\nn\\
  &+& e_q^2 \tan^4\theta_{\rm W}
  \log{Q^2 +M_{\rm Z}^2\over M_{\rm Z}^2}
\nn\\
  &+& \left. 2e_q^2
  \tan^2\theta_{\rm W}
  \log{Q^2 +M_{\rm Z}^2\over M_{\rm Z}^2}
\right\} \,,\\
%
  P^{{\rm e}^-_-}_{q_-}(z,Q^2) &=&  {3\alpha \over 2\pi} \Phi_+(z) \left\{
  e_q^2
  \log {Q^2\over m_e^2}
  \right.
\nn\\
  &+& z_q^2 \rW^2 \tan^4\theta_{\rm W}
  \log{Q^2 +M_{\rm Z}^2\over M_{\rm Z}^2}
\nn\\
  &+& 2e_q z_q \rW \tan^2\theta_{\rm W}
  \log{Q^2 +M_{\rm Z}^2\over M_{\rm Z}^2}
\nn\\
  &+& \left. (1+\rW)^2 \delta_{q \rm d}  
  \log{Q^2 +M_{\rm W}^2\over M_{\rm W}^2}
\right\} \,,
\\
%
  P^{{\rm e}^-_+}_{q_-}(z,Q^2) &=&  {3\alpha \over 2\pi} \Phi_-(z) \left\{
  e_q^2
  \log {Q^2\over m_e^2}
  \right.
\nn\\
  &+& z_q^2 \tan^4\theta_{\rm W}
  \log{Q^2 +M_{\rm Z}^2\over M_{\rm Z}^2}
\nn\\
  &-&\left. 2e_q z_q \tan^2\theta_{\rm W}
  \log{Q^2 +M_{\rm Z}^2\over M_{\rm Z}^2}
\right\} \,,
\eeqal
where $\Phi_-(z)$ and $\rW$ are defined by \Eq{Phi-def} and \Eq{zq-def},
respectively.
The splitting functions for an antiquark of the opposite helicity can be
obtained from \Eq{spfpol} by interchanging $\Phi_+$ with $\Phi_-$.

\end{document}